\documentclass[manuscript=article,journal=apchd5,layout=twocolumn]{achemso}

\setkeys{acs}{articletitle = true}

\usepackage{amsmath}
\usepackage[utf8]{inputenc}
\usepackage{graphicx}
\graphicspath{ {figure/} }
\usepackage{hyperref}

\usepackage{gensymb}

\author{F. Pisani}
\affiliation[university of Pisa]
{Enrico Fermi department of Physics, University of Pisa, 56127 Pisa, Italy}
\email{f.pisani3@studenti.unipi.it}
\author{L. Fedeli}
\affiliation[university of Milan]
{Department of Energy, Politecnico di Milano, 20133 Milano, Italy}
\email{luca.fedeli@polimi.it}
\author{A. Macchi}
\affiliation[INO]{National Institute of Optics, National Research Council (CNR/INO) A.Gozzini unit , 56124 Pisa, Italy}
\altaffiliation{Enrico Fermi department of Physics, University of Pisa, 56127 Pisa, Italy}
\email{andrea.macchi@ino.cnr.it}

\keywords{Ultrashort Pulses; Femtosecond Plasmonics; Grating Coupler; Few-cycle surface plasmon polaritons; Rotating Wavefront Pulses}

\title{Few-cycle Surface Plasmon Polariton Generation by Rotating Wavefront Pulses}

\begin{document}

\begin{abstract}
A concept for the efficient generation of surface plasmon polaritons (SPPs) with a duration of very few cycles is presented. The scheme is based on grating coupling and laser pulses with wavefront rotation (WFR), so that the resonance condition for SPP excitation is satisfied only for a time window shorter than the driving pulse. The feasibility and robustness of the technique is investigated by means of simulations with realistic parameters. In optimal conditions, we find that a $29.5$~fs pulse with $800$~nm wavelength can excite a $3.8$~fs SPP ($\sim 1.4$ laser cycles) with a peak field amplitude $2.7$ times the peak value for the laser pulse.
\end{abstract}

\section*{Introduction}

Ultrafast plasmonics raised a great interest in the last decades. Ultrashort surface plasmon polaritons (SPPs) can be employed in ultrafast photocathodes\cite{li2013,polyakov2013} to study phenomena with short time scale, e.g. the effects of the vibration of a lattice on the electric and thermal transport\cite{temnov2013}, molecular chemical processes\cite{scrinzi2005} and the electrons motion in an atom. The possibility to manipulate the SPPs with plasmonic switches and modulators\cite{macdonald2010} makes them powerful instruments for the next generation of photonic circuits\cite{gramotnev2010,wei2012,rewitz2011}. Moreover, few-cycles SPPs show a strong modulation of the fields oscillation over a laser period. This significant asymmetry and its dependency on the carrier envelope phase (CEP) of the laser pulse is important to measure phase related phenomena as photo-electron emission\cite{paulus2003} and for CEP detectors\cite{racz2011,kruger2011,piglosiewicz2014}. Molecular sensing\cite{jain2008} and spectroscopy techniques based on SPPs, such as SERS\cite{kumar2012,keller2015} have also found widespread diffusion in the material science community. Combined with the SPP potential to concentrate the light beyond the diffraction limit and to increase the energy transfered to a material, the ultrashort duration makes SPPs well suitable for energy concentration\cite{barnes2003} also in view of high intensity applications such as generation of short pulses of high-energy electrons\cite{Fedelielectronaccelerator} and high-order harmonics of the laser light\cite{fedeli2016}. For such applications a technique to produce SPPs with few cycle duration and high intensity may be useful.

In this paper we present a concept to generate ultrashort SPPs with a duration of less than two laser cycles. The proposed method is based on laser pulses with wavefront rotation (WFR) impinging on a metallic grating. The use of a grating \cite{ye2014} is a well known method to couple an electromagnetic (EM) wave to a SPP. When the plasma frequency of the grating material is much greater than the laser frequency the condition for resonant SPP excitation \cite{Maier} is satisfied for values $\theta_R$ of the incidence angle $\theta$ which are solutions of the equation
\begin{equation}
\sin\theta\simeq1\pm n\frac{\lambda}{d},
\label{solution}
\end{equation} 
where $\lambda$ is the laser wavelength, $d$ is the grating pitch and $n$ is an integer number (in the following we restrict to the $|n|=1$ case). 
In a laser pulse with WFR, the resonance condition will be satisfied only for a short time interval since the rotation of the fronts of constant phase is equivalent to a continuous temporal variation of the local angle of incidence. As a consequence, the generation of a SPP with a duration much shorter than the impinging laser pulse is expected.

A pulse with WFR can be generated by focusing a pulse with the wavefronts tilted with respect to the direction of propagation \cite{Quere,wheeler2012}. Heuristically, different portions of the laser pulse reach the focus at different times and with different directions of propagation, thus a pulse with wavefronts rotating in time is observed at the focus.
The expression of a Gaussian pulse with WFR at focus is (see \cite{Vincenti,Akturk} for a more detailed description):
\begin{align}
E(r,t)&=E_0\exp\Big[-\frac{r^2}{w_0^2}-\frac{t^2}{\tau^2}\Big]\exp\Big[i\varphi(r,t)\Big],\\
\varphi(r,t)&=4\frac{w_i\eta}{w_0\tau_f\tau_i}rt+\omega_Lt,
     \end{align}
where $\omega_L$ is the laser frequency, $\tau$, $w_0$, $r$ are respectively the pulse duration, waist and transverse coordinate in the focus, $\tau_i$ and $w_i$ are the duration and the waist before focusing, and $\eta$ is the pulse front tilt parameter. Notice that the pulse is characterized by spatio-temporal coupling, i.e. in the expression of the EM field the spatial and temporal dependences cannot be separated  [$E({\bf r},t)\neq E_0S({\bf r})T(t)$].
The rotation velocity of the phase front is given by 
\begin{equation}
v_r=\frac{c}{\omega_L}\frac{\partial^2\varphi(r,t)}{\partial r\partial t }
\end{equation}
In their experiment, Quéré et al. \cite{Quere} demonstrated the possibility to control the rotation velocity and obtained a value of $v_r\simeq14.6$~mrad/fs. A maximum achievable value of $\sim30$ mrad/fs was also estimated with their setup, so that a $30$ fs laser pulse would span a $\sim50\degree$~arc. 

We performed a two-dimensional (2D) simulation campaign to test the effectiveness and robustness of the proposed concept. As it will be shown below, the simulations demonstrate that a SPP as short as $1.4$ laser cycles may be generated, with a peak value of the field $2.7$ times the value for the driving pulse and an energy conversion efficiency of $5.6\%$.

\subsection*{Result and Discussion}

In the simulations using the MEEP code (see Methods), Silver was selected as the grating material, which is a typical choice in plasmonics \cite{smith2015,gu2012,jain2008,barnes2003}. The linear response is described by the dielectric function
\begin{equation}
\varepsilon(\omega)=1-\frac{\omega_P^2}{\omega^2-i\gamma\omega} \; ,
\end{equation}
with the plasma frequency $\omega_P=9.6$~eV ($6.8\times 10^{14}$~s$^{-1}$) and the damping coefficient $\gamma=0.0228$~eV (values from \cite{blaber2009}). The wavelength of the laser pulse was $\lambda=800$~nm (Ti:Sapphire). While we restrict to a single material and a single laser source for simplicity, our concept does not depend on this particular choice.

\begin{figure}[t!]
    \includegraphics[width=0.45\textwidth]{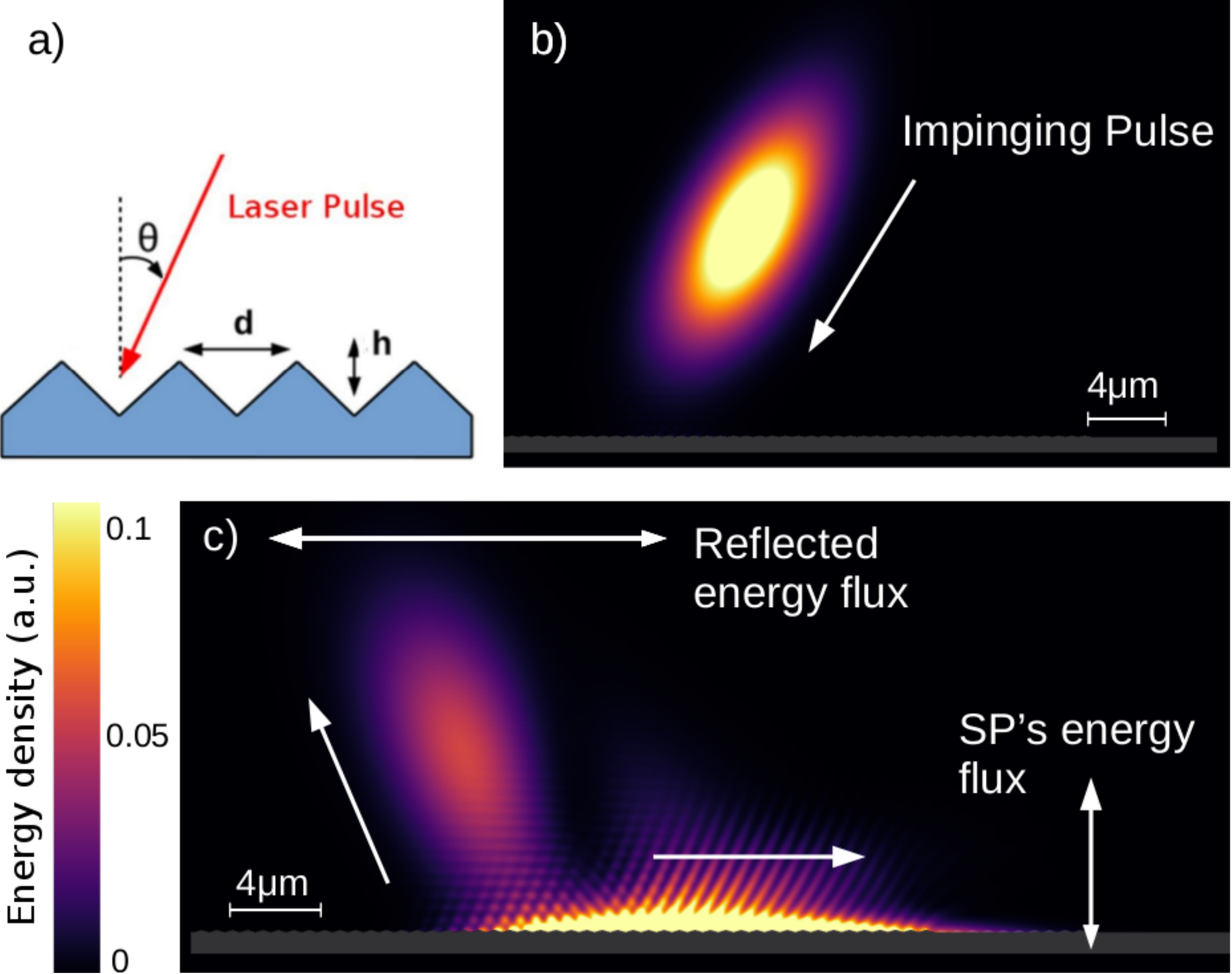}
    \caption{a): Schematic setup of the simulation box. b): 2D map of the EM energy density (cycle-averaged) showing  a pulse without WFR impinging on the grating (shown in grey) is reported. c): Same as~b) at later times, showing the excited SPP propagating along the grating surface and the reflected pulse. The associated energy fluxes are measured through the marked vertical and horizontal lines.}
    \label{box}
\end{figure}

In Fig.\ref{box}~a) the basic simulation setup is shown. The laser pulse impinges in the middle of the grating coupler, which is $\sim25\mu$m wide. On the right side of the target, where the SPP propagates, the surface is flat.  
Fig.\ref{box}~b) shows the snapshot of a laser pulse (without WFR) impinging on the grating at the resonant angle. Fig.\ref{box}~c) shows the excited SPP and the reflected pulse. 
The energy fluxes of the reflected pulse and SPP are measured through the marked vertical and horizontal lines. The laser focus was at a distance of $15\mu$m from the target. 

In order to optimize the coupling efficiency a preliminary study has been performed by varying both the parameters of the laser pulse and of the grating, i.e. the depth and width of the grooves and the profile shape (triangular, rectangular, sinusoidal and saw-tooth like). The efficiency was maximal for a triangular grating having depth $h=140$~nm and pitch $d=0.56$ $\mu$m, corresponding to the resonance angle $\theta_R=-25\degree$ (The negative value of the incidence angle corresponds to the SPP propagating in the opposite direction with respect to $k_{||}$, the component of the laser pulse wavevector parallel to the surface). The maximum of the coupled energy (i.e. the percentage of the pulse energy transferred to the material) was $61\%$ of that of the laser pulse. The energy transferred to the SPP, measured after $\sim30$~fs from the generation and after a propagation of $\sim12~\mu$m along the surface, was $38\%$. The remaining fraction of the coupled energy was either re-emitted during the propagation along the grating region or absorbed in the material. 

\begin{figure}[b!]
    \includegraphics[width=0.45\textwidth]{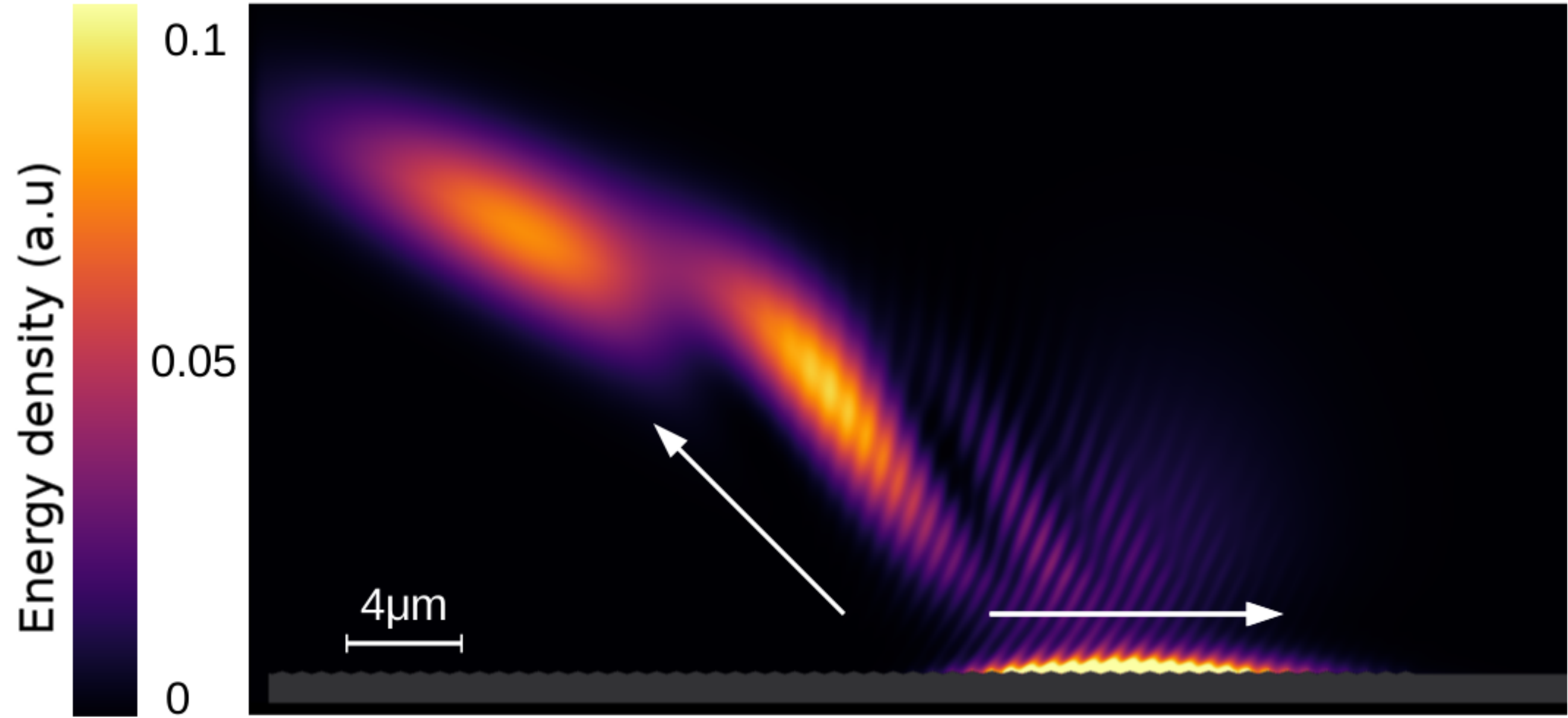}
    \caption{EM energy density after the interaction of a WFR pulse with the grating (other pulse parameters are as in Fig.\ref{box}. A SPP much shorter than the laser pulse is generated. The reflected pulse profile shows a narrow minimum (``hole'') corresponding to the SPP excitation near the pulse peak. The laser pulse had a duration of $29.5$ fs, a waist of $4\lambda$ and a rotation parameter $\xi=0.2$.
    }
    \label{hole}
\end{figure}

Fig.\ref{hole} shows a snapshot of the field for the same parameters of Fig.\ref{box} but using a pulse with WFR. The rotation of the wavefronts has been set in order to have the central wavefront (i.e. the phase front corresponding to the maximum intensity) impinging at an angle $90\degree-\theta_R$ on the target, so that the instantaneous angle of incidence has the resonant value at the pulse peak.
A SPP shorter than the laser pulse is generated near the peak of the pulse. In coincidence to the SPP excitation, a narrow ``hole'' is produced in the profile of the reflected pulse corresponding to the transient decrease in the reflectivity of the grating. 

\begin{figure}[t]
    \includegraphics[width=0.45\textwidth]{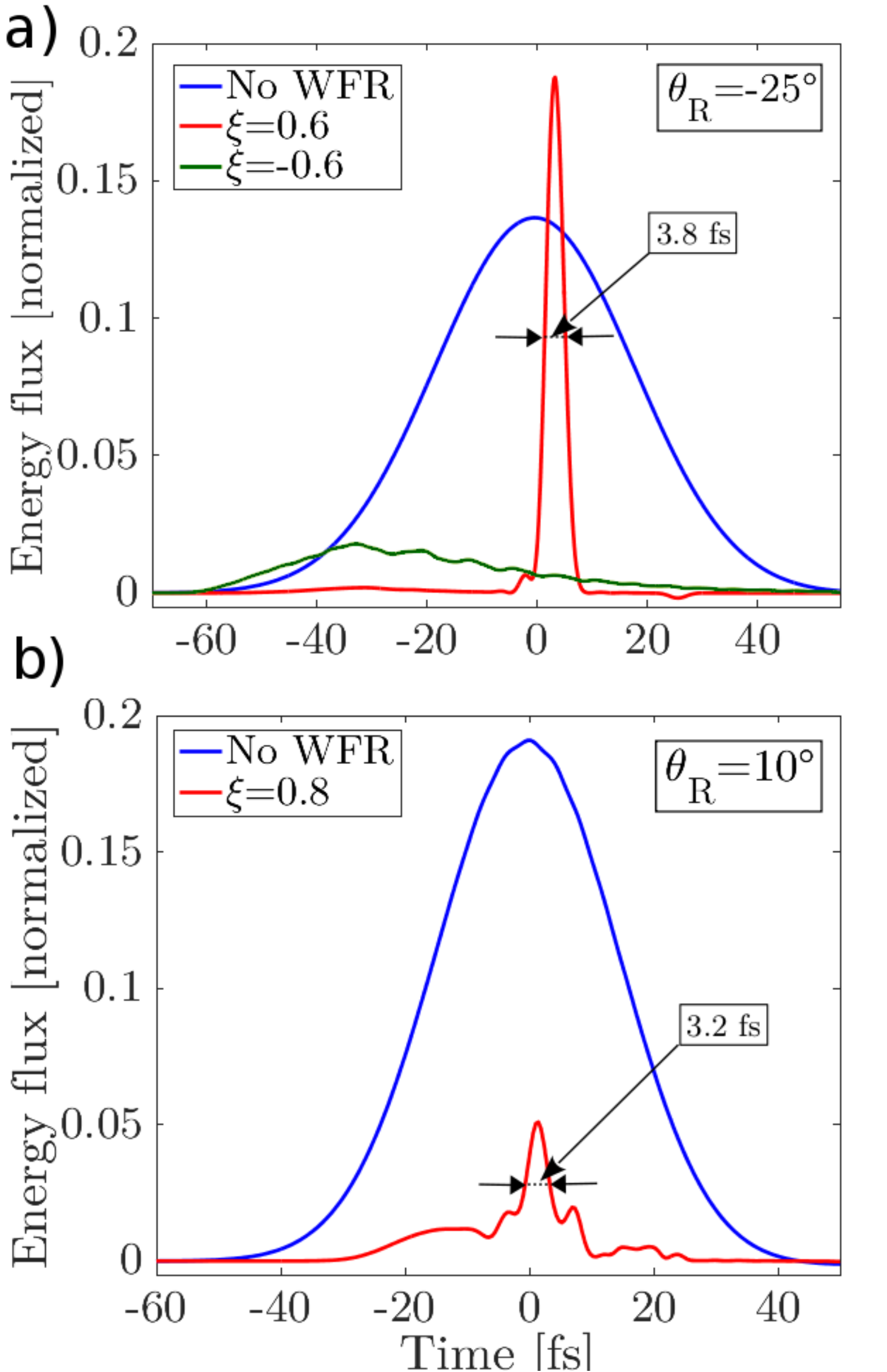}
    \caption{a)~Energy fluxes (normalized to the energy flux of the laser pulse) of SPPs generated on a grating by laser pulses incident at the resonant angle $\theta_R=-25\degree$ and different WFR velocities $\xi=0$ (blue curve), $\xi=0.6$ (red curve) and $\xi=-0.6$ (green curve). The laser pulse duration and waist size are $\sim 29.5$~fs and $4.8\mu$m, respectively.
b)~Same as a) but for $\theta_R=10\degree$ and $\xi=0$ (blue curve) and $\xi=0.8$ (red curve).}
    \label{fluxes}
\end{figure}

Fig.\ref{fluxes}~a) shows the energy flux of the SPPs generated on the grating with pitch $d=0.56$ $\mu$m ($\theta_R=-25\degree$) by a WFR  pulse  with a duration of $\sim 29.5$~fs and a waist size $6\lambda=4.8\mu$m.  Using a rotation parameter $\xi=0.6$, the SPP has a duration of $\sim3.8$~fs, i.e. $1.4$ laser cycles at full-width-half-maximum (FWHM). The energy coupling efficiency is $5.6\%$ and the peak amplitude of the SPP field is $\sim2.7$ times the peak amplitude of the laser pulse. The intensity of the ultrashort SPP is also higher (by  $\sim 35\%$) than the intensity of the SPP generated without WFR, which has almost the same duration as the laser pulse. This intensity increase (which will be further discussed below) shows a clear advantage of the WFR approach with respect to the alternative idea of exploiting the laser bandwidth to produce a chirped pulse, so that the laser frequency is resonant with that of the SPP only for a portion of the pulse. In this case, the driving pulse is stretched in time and has a lower intensity, resulting in a low efficiency and modest shortening effect as we verified via our simulations.

\begin{figure}[t]
    \includegraphics[width=0.45 \textwidth]{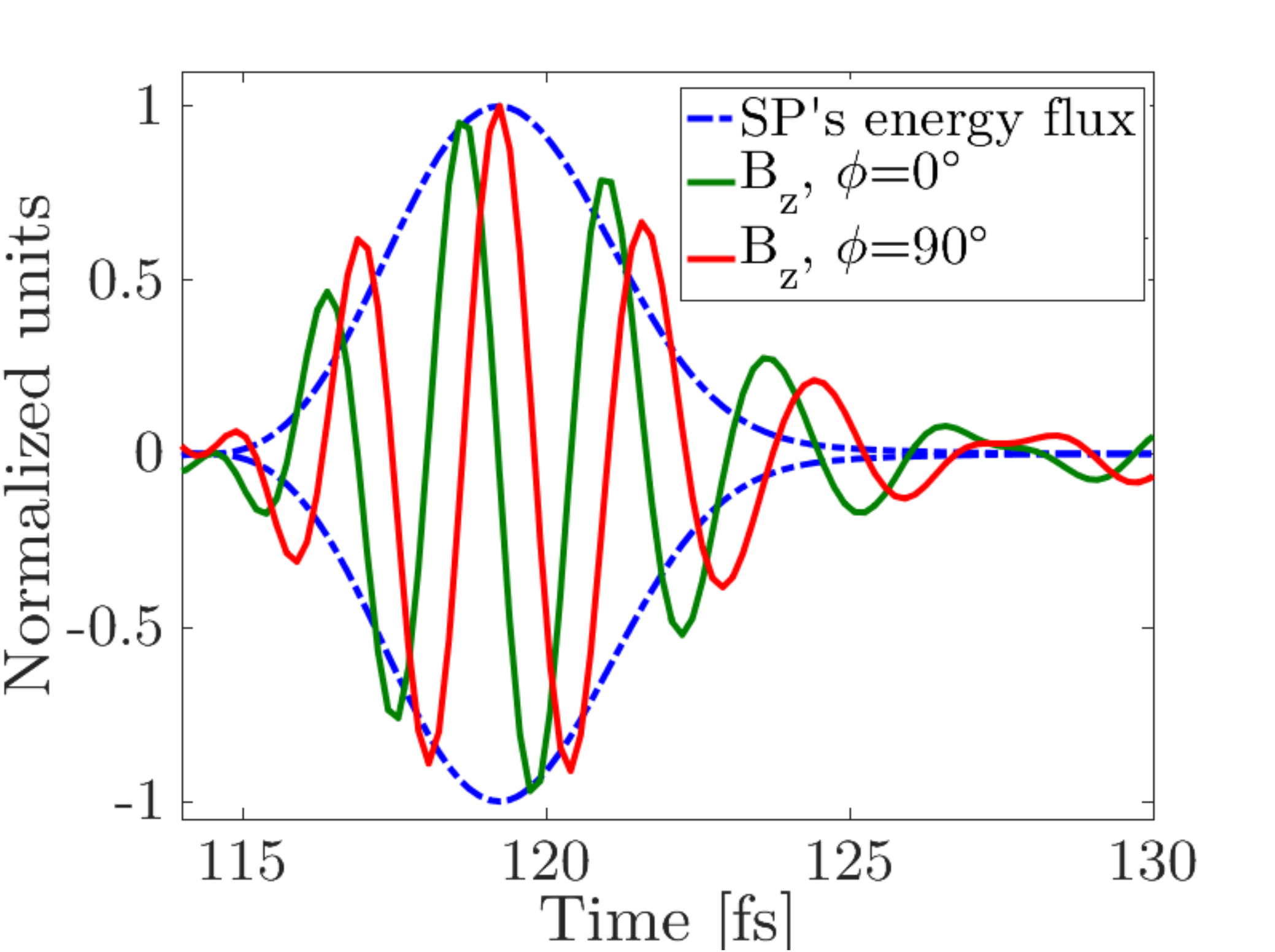}
    \caption{Energy flux (dotted line) and magnetic field (continuous lines) as a function of time for two SPPs of $\sim3.8$~fs duration generated by two WFR laser pulses with $\xi=0.6$ and different values of the CEP $\phi$ ($0\degree$ and $90\degree$). Other parameters are the same as in Fig.\ref{fluxes}~a).   
The SPP generated for $\phi=90\degree$ show a significant asymmetry in the waveform. 
}
    \label{shortest}
\end{figure}

The duration of the SPP is short enough for the absolute phase of the field oscillation to be important for the SPP characterization.
Fig.\ref{shortest} shows the energy flux and the magnetic field as functions of time for the same ultrashort SPP of Fig.\ref{fluxes}~a) and a SPP generated in identical conditions but for the value of the carrier envelope phase (CEP) $\phi$ of the laser pulse, which differs by $90\degree$. The two values of $\phi$ yield a SPP with zero and $11\%$ asymmetry, defined as the relative difference between the absolute values of the maximum and the minumum of the field.

Using extremely high values of $\xi$ is found to affect the envelope of the SPPs. As an example Fig.\ref{fluxes}~b) shows a case analogous to Fig.\ref{fluxes}~a) for $\theta_r=10\degree$ ($d=0.98$ $\mu$m), in which the effect of $\xi=0.8$ is compared to $\xi=0$. The SPP shows significant modulations with a central peak having duration of $\sim3.2$ fs ($1.2$ laser cycles) and energy flux characterized by several secondary peaks with an intensity comparable to that of the main one. Such modulated SPP might not be suitable for all the foreseen applications of ultrashort plasmonics. 
Notice that the direction of the WFR, i.e the sign of the rotation parameter $\xi$, is crucial for the ultrashort SPP generation. As also shown in Fig.\ref{fluxes}~a), the SPP generated with $\xi=-0.6$ has a FWHM duration of $39.6$~fs, i.e. longer than that of the driving pulse, and a very low intensity. 
This effect appears to be related to the pulse impinging in different points of the grating because of the WFR. Different ``portions'' of the pulse propagate in different directions and the focus of the pulse is not exactly on the target surface, thus the incidence point (determined by the position of the maximum of the energy flux on the grating surface) is not fixed but moves in time. Depending on the sign of the rotation velocity, the incidence point can move either in the direction of propagation of the SPP or in the opposite direction. In the first case, the laser pulse ``follows'' the SPP and can sustain its growth, which produces the observed increase of the peak intensity with respect to the SPP excited without WFR.   
In order to give a further illustration of the effect of the sign of the rotation velocity, in the Supplementary Material we show simulations for the ``resonant'' case at normal laser incidence $\theta=0\degree$, corresponding to $d=\lambda$. In the case of a pulse without WFR, in this configuration two counter propagating SPPs of same amplitude and duration are excited. When adding WFR, only the SPP which propagates in the same direction as the incidence point is generated with high efficiency. 

\begin{figure*}[]
\includegraphics[width=0.95\textwidth]{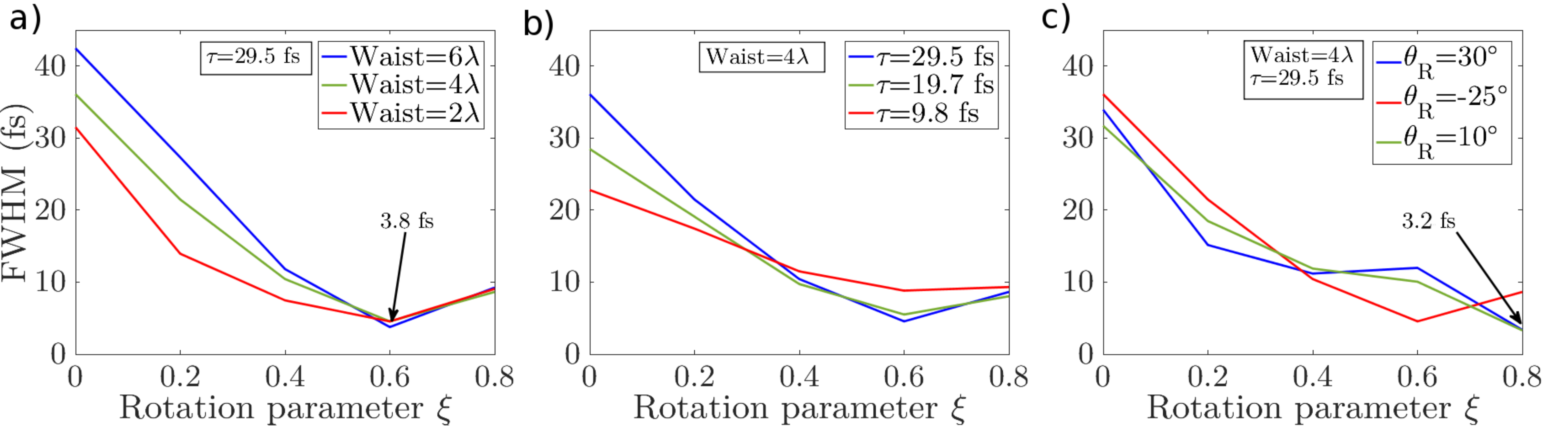}
    \caption{Duration of the SPP at FWHM as a function of the WFR parameter $\xi$ and for different values of the a) laser pulse waist, b) laser pulse duration, c) grating resonant angle.}
    \label{allcompared}
\end{figure*}

In order to test the robustness of the proposed technique and to check the sensitivity of the ultrashort SPP to laser and grating characteristics, 
a parametric study has been performed measuring the duration of the SPP as a function of the rotation velocity $\xi$ for different values of the laser pulse waist and duration and of the grating pitches i.e. of the resonant angle $\theta_R$. In these parametric simulations, all parameters but one are identical to the case of the ultrashort SPP with 1.4~cycle duration in Fig.\ref{fluxes} and Fig.{shortest}. 
Results are summarized in Fig.\ref{allcompared}. 
For small values of $\xi$, the SPP duration increases with the waist size $w_0$ (Fig.\ref{allcompared}~a). This is due to a geometrical effect, since the time needed to propagate across the lit zone, over which the SPP is excited , is not negligible with respect to the laser pulse duration.  
However, for the ``optimal value'' $\xi=0.6$ the SPP duration is almost independent on $w_0$. Reducing the laser pulse duration also has no effect on the shortest SSP duration (Fig.\ref{allcompared}~b). Some dependence on the resonant angle $\theta_R$ is apparent (Fig.\ref{allcompared}~c). The case $\theta_R=25\degree$ appears to be the most suitable since the shortest SPP duration is obtained for not so ``extreme'' values of $\xi$.

\begin{figure}[]
    \includegraphics[width=0.45 \textwidth]{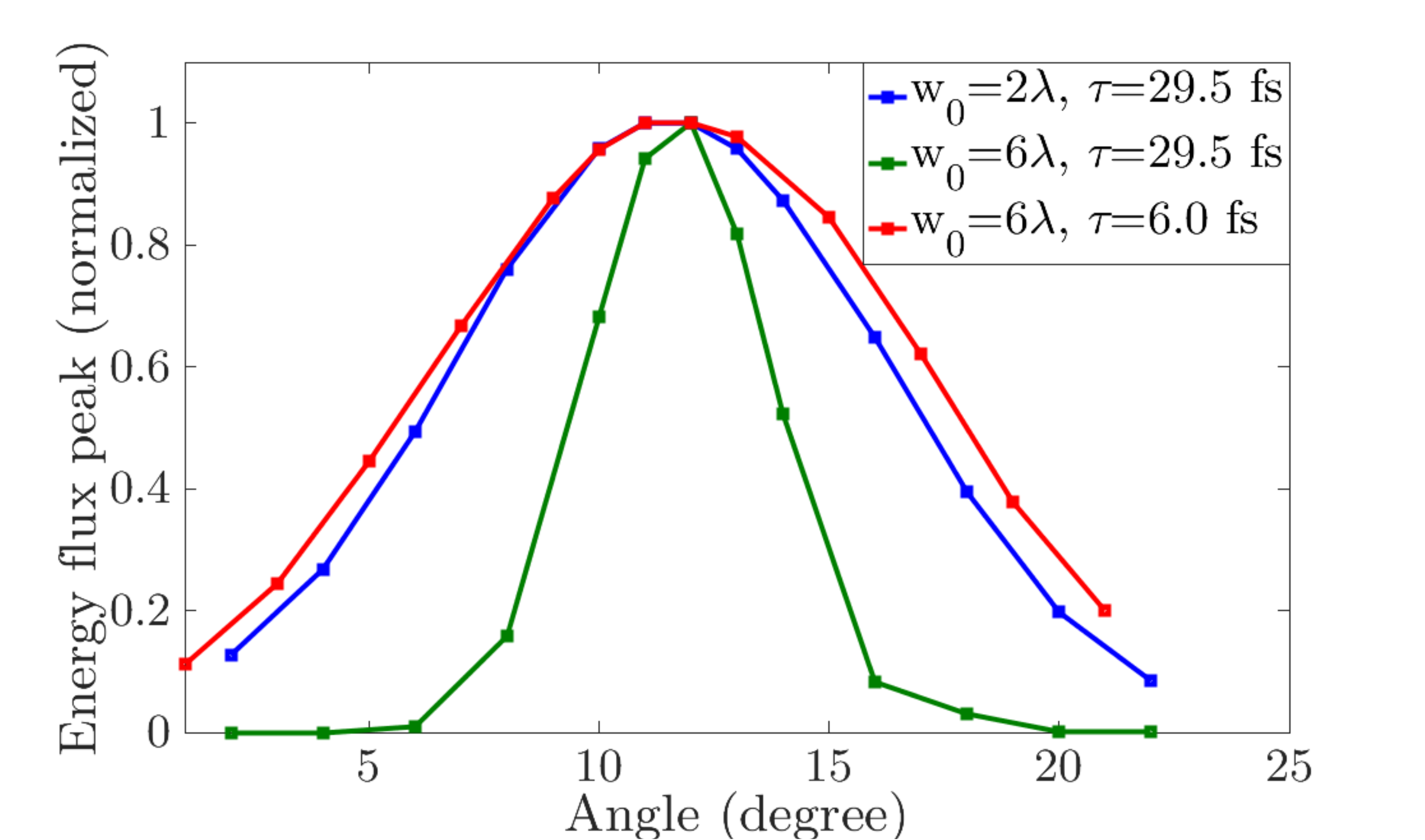}
    \caption{SPP's energy flux peak as a function of the incidence angle for a pulse with waist $2\lambda$ and $6\lambda$ with a duration of $29.5$ fs and for a pulse with waist $6\lambda$ and a duration of $6.0$ fs. The simulations were performed with a grating having a pitch $d=0.98$ $\mu$m, corresponding to a resonance angle of $\sim10\degree$.}
    \label{reswaist}
\end{figure}
Finally, the energy flux of the SPP has been measured as a function of the incidence angle $\theta$, to characterize the width in $\theta$ of the SPP resonance. Results are shown in Fig.\ref{reswaist}. The resonance is tighter for a pulse with waist $w_0=6\lambda$ ($\Delta\theta_R\simeq 4.7\degree$) with respect to a pulse with shorter waist $w_0=2\lambda$ ($\Delta\theta_R\simeq10.9\degree$). This is due to the spread in the wavevector spectrum for a focused pulse, which is of the order of $\lambda/w_0$. The pulse bandwidth also contributes to the resonance width because of the relation between $\theta_R$ and $\lambda$. As also shown in Fig.\ref{reswaist}, the reference pulse of $\sim29.5$~fs duration and bandwidth of $\sim23 $~THz leads to a tighter resonance than using a pulse of $6.0$~fs duration and $\sim105$ THz bandwidth, for which $\Delta\theta_R\simeq12.5\degree$. 

\section*{Conclusion}

A proposal for the generation of ultrashort surface plasmon polaritons (SPPs) by laser pulses with wavefront rotation (WFR) has been investigated via simulations with realistic parameters. Results show that the duration of a SPP can be controlled with the rotation velocity and a SPP of $\sim3.8$~fs duration, corresponding to $1.4$ optical cycles of the $800$~nm driving laser, can be generated with high amplitude ($\sim2.7$ times the laser field) and coupling efficiency ($<5\%$). The SPP is so short that carrier-envelope phase effects become important. The experimental realization appears to be straightforward, as it could employ already tested schemes for WFR\cite{Quere} and characterize the SPP duration via, e.g., attosecond photoscopy \cite{lupetti2014}. The few-cycle, high intensity SPPs may find wide application in ultrafast plasmonics, such as for e.g. ultra-fast surface enhanced Raman spectroscopy \cite{keller2015} or plasmonic circuitry \cite{gramotnev2010,wei2012,rewitz2011}. 

 \section*{Methods}
The simulation campaign was performed with a finite difference time domain code, MEEP \cite{Oskooi2009}.
The expression for the WFR laser pulse in the focus was:
\begin{equation}
E_{WFR}=E_G(r,z,t)\exp\Big[i\omega_Lt+itr\xi\Big],
\end{equation}
where $\displaystyle\xi=4\frac{w_i\eta}{w_0\tau_f\tau_i}$ is the rotation parameter. $E_G(r,z,t)$ is the expression that describes the spatial profile of a 2D Gaussian pulse \cite{Siegman}:
\begin{align}
    E_{G}(r,z,t) & =\Big(\frac{2w_0^2}{\pi w(z)^2}\Big)^{1/4}\exp\Big[-\frac{r^2}{w(z)^2} -\frac{t^2}{\tau^2}\Big] \nonumber \\
   & \exp\Big[-ikz -ik\frac{r^2}{2R(z)} +i\psi(z)\Big]\,,
\end{align}
where $z$ and $r$ represent respectively the longitudinal and transverse coordinates, $w_0$ and $\tau$ are the pulse waist and duration and $z_R$ is the Rayleigh length. $\psi(z)$, $R(z)$ and $w(z)$ are the Gouy phase, the radius of curvature of the  wavefronts and the diameter of the pulse\cite{Siegman}:
\begin{equation}
    R(z) =z\Big[1+\Big(\frac{z_R}{z}\Big)^2\Big]\,,
\end{equation}
\begin{equation}
    w(z) =w_0\sqrt{1+\Big(\frac{z_R}{z}\Big)^2}\,,
\end{equation}
\begin{equation}
    \phi(z) =\arctan\Big(\frac{z}{z_R}\Big)\,.
\end{equation}

The rotation velocity was controlled varying the rotation parameter $\xi$. The wavefronts rotate in time with a rotation velocity $v_r$ determined by $v_r(\xi)\simeq37.7\xi$ mrad/fs, so that the maximum value estimated by Quéré \cite{Quere} correspond to $\xi\simeq0.8$ in our simulations. The box size was $40\times30$ $\mu$m with a resolution of $10$ nm.

%\bibliography{bibliography}

\begin{mcitethebibliography}{31}
\providecommand*\natexlab[1]{#1}
\providecommand*\mciteSetBstSublistMode[1]{}
\providecommand*\mciteSetBstMaxWidthForm[2]{}
\providecommand*\mciteBstWouldAddEndPuncttrue
  {\def\EndOfBibitem{\unskip.}}
\providecommand*\mciteBstWouldAddEndPunctfalse
  {\let\EndOfBibitem\relax}
\providecommand*\mciteSetBstMidEndSepPunct[3]{}
\providecommand*\mciteSetBstSublistLabelBeginEnd[3]{}
\providecommand*\EndOfBibitem{}
\mciteSetBstSublistMode{f}
\mciteSetBstMaxWidthForm{subitem}{(\alph{mcitesubitemcount})}
\mciteSetBstSublistLabelBeginEnd
  {\mcitemaxwidthsubitemform\space}
  {\relax}
  {\relax}

\bibitem[Li \latin{et~al.}(2013)Li, To, Andonian, Feng, Polyakov, Scoby,
  Thompson, Wan, Padmore, and Musumeci]{li2013}
Li,~R.; To,~H.; Andonian,~G.; Feng,~J.; Polyakov,~A.; Scoby,~C.; Thompson,~K.;
  Wan,~W.; Padmore,~H.; Musumeci,~P. Surface-plasmon resonance-enhanced
  multiphoton emission of high-brightness electron beams from a nanostructured
  copper cathode. \emph{Physical review letters} \textbf{2013}, \emph{110},
  074801\relax
\mciteBstWouldAddEndPuncttrue
\mciteSetBstMidEndSepPunct{\mcitedefaultmidpunct}
{\mcitedefaultendpunct}{\mcitedefaultseppunct}\relax
\EndOfBibitem
\bibitem[Polyakov \latin{et~al.}(2013)Polyakov, Senft, Thompson, Feng, Cabrini,
  Schuck, Padmore, Peppernick, and Hess]{polyakov2013}
Polyakov,~A.; Senft,~C.; Thompson,~K.; Feng,~J.; Cabrini,~S.; Schuck,~P.;
  Padmore,~H.; Peppernick,~S.~J.; Hess,~W.~P. Plasmon-enhanced photocathode for
  high brightness and high repetition rate x-ray sources. \emph{Physical review
  letters} \textbf{2013}, \emph{110}, 076802\relax
\mciteBstWouldAddEndPuncttrue
\mciteSetBstMidEndSepPunct{\mcitedefaultmidpunct}
{\mcitedefaultendpunct}{\mcitedefaultseppunct}\relax
\EndOfBibitem
\bibitem[Temnov \latin{et~al.}(2013)Temnov, Klieber, Nelson, Thomay, Knittel,
  Leitenstorfer, Makarov, Albrecht, and Bratschitsch]{temnov2013}
Temnov,~V.~V.; Klieber,~C.; Nelson,~K.~A.; Thomay,~T.; Knittel,~V.;
  Leitenstorfer,~A.; Makarov,~D.; Albrecht,~M.; Bratschitsch,~R. Femtosecond
  nonlinear ultrasonics in gold probed with ultrashort surface plasmons.
  \emph{Nature communications} \textbf{2013}, \emph{4}, 1468\relax
\mciteBstWouldAddEndPuncttrue
\mciteSetBstMidEndSepPunct{\mcitedefaultmidpunct}
{\mcitedefaultendpunct}{\mcitedefaultseppunct}\relax
\EndOfBibitem
\bibitem[Scrinzi \latin{et~al.}(2005)Scrinzi, Ivanov, Kienberger, and
  Villeneuve]{scrinzi2005}
Scrinzi,~A.; Ivanov,~M.~Y.; Kienberger,~R.; Villeneuve,~D.~M. Attosecond
  physics. \emph{Journal of Physics B: Atomic, Molecular and Optical Physics}
  \textbf{2005}, \emph{39}, R1\relax
\mciteBstWouldAddEndPuncttrue
\mciteSetBstMidEndSepPunct{\mcitedefaultmidpunct}
{\mcitedefaultendpunct}{\mcitedefaultseppunct}\relax
\EndOfBibitem
\bibitem[MacDonald and Zheludev(2010)MacDonald, and Zheludev]{macdonald2010}
MacDonald,~K.~F.; Zheludev,~N.~I. Active plasmonics: current status.
  \emph{Laser \& Photonics Reviews} \textbf{2010}, \emph{4}, 562--567\relax
\mciteBstWouldAddEndPuncttrue
\mciteSetBstMidEndSepPunct{\mcitedefaultmidpunct}
{\mcitedefaultendpunct}{\mcitedefaultseppunct}\relax
\EndOfBibitem
\bibitem[Gramotnev and Bozhevolnyi(2010)Gramotnev, and
  Bozhevolnyi]{gramotnev2010}
Gramotnev,~D.~K.; Bozhevolnyi,~S.~I. Plasmonics beyond the diffraction limit.
  \emph{Nature photonics} \textbf{2010}, \emph{4}, 83--91\relax
\mciteBstWouldAddEndPuncttrue
\mciteSetBstMidEndSepPunct{\mcitedefaultmidpunct}
{\mcitedefaultendpunct}{\mcitedefaultseppunct}\relax
\EndOfBibitem
\bibitem[Wei and Xu(2012)Wei, and Xu]{wei2012}
Wei,~H.; Xu,~H. Nanowire-based plasmonic waveguides and devices for integrated
  nanophotonic circuits. \emph{Nanophotonics} \textbf{2012}, \emph{1},
  155--169\relax
\mciteBstWouldAddEndPuncttrue
\mciteSetBstMidEndSepPunct{\mcitedefaultmidpunct}
{\mcitedefaultendpunct}{\mcitedefaultseppunct}\relax
\EndOfBibitem
\bibitem[Rewitz \latin{et~al.}(2011)Rewitz, Keitzl, Tuchscherer, Huang,
  Geisler, Razinskas, Hecht, and Brixner]{rewitz2011}
Rewitz,~C.; Keitzl,~T.; Tuchscherer,~P.; Huang,~J.-S.; Geisler,~P.;
  Razinskas,~G.; Hecht,~B.; Brixner,~T. Ultrafast plasmon propagation in
  nanowires characterized by far-field spectral interferometry. \emph{Nano
  letters} \textbf{2011}, \emph{12}, 45--49\relax
\mciteBstWouldAddEndPuncttrue
\mciteSetBstMidEndSepPunct{\mcitedefaultmidpunct}
{\mcitedefaultendpunct}{\mcitedefaultseppunct}\relax
\EndOfBibitem
\bibitem[Paulus \latin{et~al.}(2003)Paulus, Lindner, Walther, Baltu{\v{s}}ka,
  Goulielmakis, Lezius, and Krausz]{paulus2003}
Paulus,~G.~G.; Lindner,~F.; Walther,~H.; Baltu{\v{s}}ka,~A.; Goulielmakis,~E.;
  Lezius,~M.; Krausz,~F. Measurement of the phase of few-cycle laser pulses.
  \emph{Physical review letters} \textbf{2003}, \emph{91}, 253004\relax
\mciteBstWouldAddEndPuncttrue
\mciteSetBstMidEndSepPunct{\mcitedefaultmidpunct}
{\mcitedefaultendpunct}{\mcitedefaultseppunct}\relax
\EndOfBibitem
\bibitem[R{\'a}cz \latin{et~al.}(2011)R{\'a}cz, Irvine, Lenner, Mitrofanov,
  Baltu{\v{s}}ka, Elezzabi, and Dombi]{racz2011}
R{\'a}cz,~P.; Irvine,~S.; Lenner,~M.; Mitrofanov,~A.; Baltu{\v{s}}ka,~A.;
  Elezzabi,~A.; Dombi,~P. Strong-field plasmonic electron acceleration with
  few-cycle, phase-stabilized laser pulses. \emph{Applied Physics Letters}
  \textbf{2011}, \emph{98}, 111116\relax
\mciteBstWouldAddEndPuncttrue
\mciteSetBstMidEndSepPunct{\mcitedefaultmidpunct}
{\mcitedefaultendpunct}{\mcitedefaultseppunct}\relax
\EndOfBibitem
\bibitem[Kr{\"u}ger \latin{et~al.}(2011)Kr{\"u}ger, Schenk, and
  Hommelhoff]{kruger2011}
Kr{\"u}ger,~M.; Schenk,~M.; Hommelhoff,~P. Attosecond control of electrons
  emitted from a nanoscale metal tip. \emph{Nature} \textbf{2011}, \emph{475},
  78--81\relax
\mciteBstWouldAddEndPuncttrue
\mciteSetBstMidEndSepPunct{\mcitedefaultmidpunct}
{\mcitedefaultendpunct}{\mcitedefaultseppunct}\relax
\EndOfBibitem
\bibitem[Piglosiewicz \latin{et~al.}(2014)Piglosiewicz, Schmidt, Park,
  Vogelsang, Gro{\ss}, Manzoni, Farinello, Cerullo, and
  Lienau]{piglosiewicz2014}
Piglosiewicz,~B.; Schmidt,~S.; Park,~D.~J.; Vogelsang,~J.; Gro{\ss},~P.;
  Manzoni,~C.; Farinello,~P.; Cerullo,~G.; Lienau,~C. Carrier-envelope phase
  effects on the strong-field photoemission of electrons from metallic
  nanostructures. \emph{Nature Photonics} \textbf{2014}, \emph{8}, 37--42\relax
\mciteBstWouldAddEndPuncttrue
\mciteSetBstMidEndSepPunct{\mcitedefaultmidpunct}
{\mcitedefaultendpunct}{\mcitedefaultseppunct}\relax
\EndOfBibitem
\bibitem[Jain \latin{et~al.}(2008)Jain, Huang, El-Sayed, and
  El-Sayed]{jain2008}
Jain,~P.~K.; Huang,~X.; El-Sayed,~I.~H.; El-Sayed,~M.~A. Noble metals on the
  nanoscale: optical and photothermal properties and some applications in
  imaging, sensing, biology, and medicine. \emph{Accounts of chemical research}
  \textbf{2008}, \emph{41}, 1578--1586\relax
\mciteBstWouldAddEndPuncttrue
\mciteSetBstMidEndSepPunct{\mcitedefaultmidpunct}
{\mcitedefaultendpunct}{\mcitedefaultseppunct}\relax
\EndOfBibitem
\bibitem[Kumar(2012)]{kumar2012}
Kumar,~G.~P. Plasmonic nano-architectures for surface enhanced Raman
  scattering: a review. \emph{Journal of Nanophotonics} \textbf{2012},
  \emph{6}, 064503--1\relax
\mciteBstWouldAddEndPuncttrue
\mciteSetBstMidEndSepPunct{\mcitedefaultmidpunct}
{\mcitedefaultendpunct}{\mcitedefaultseppunct}\relax
\EndOfBibitem
\bibitem[Keller \latin{et~al.}(2015)Keller, Brandt, Cassabaum, and
  Frontiera]{keller2015}
Keller,~E.~L.; Brandt,~N.~C.; Cassabaum,~A.~A.; Frontiera,~R.~R. Ultrafast
  surface-enhanced Raman spectroscopy. \emph{Analyst} \textbf{2015},
  \emph{140}, 4922--4931\relax
\mciteBstWouldAddEndPuncttrue
\mciteSetBstMidEndSepPunct{\mcitedefaultmidpunct}
{\mcitedefaultendpunct}{\mcitedefaultseppunct}\relax
\EndOfBibitem
\bibitem[Barnes \latin{et~al.}(2003)Barnes, Dereux, and Ebbesen]{barnes2003}
Barnes,~W.~L.; Dereux,~A.; Ebbesen,~T.~W. Surface plasmon subwavelength optics.
  \emph{Nature} \textbf{2003}, \emph{424}, 824--830\relax
\mciteBstWouldAddEndPuncttrue
\mciteSetBstMidEndSepPunct{\mcitedefaultmidpunct}
{\mcitedefaultendpunct}{\mcitedefaultseppunct}\relax
\EndOfBibitem
\bibitem[Fedeli \latin{et~al.}(2016)Fedeli, Sgattoni, Cantono, Garzella,
  R\'eau, Prencipe, Passoni, Raynaud, Kv\ifmmode \check{e}\else
  \v{e}\fi{}to\ifmmode~\check{n}\else \v{n}\fi{}, Proska, Macchi, and
  Ceccotti]{Fedelielectronaccelerator}
Fedeli,~L.; Sgattoni,~A.; Cantono,~G.; Garzella,~D.; R\'eau,~F.; Prencipe,~I.;
  Passoni,~M.; Raynaud,~M.; Kv\ifmmode \check{e}\else
  \v{e}\fi{}to\ifmmode~\check{n}\else \v{n}\fi{},~M.; Proska,~J.; Macchi,~A.;
  Ceccotti,~T. Electron Acceleration by Relativistic Surface Plasmons in
  Laser-Grating Interaction. \emph{Physical Review Letters} \textbf{2016},
  \emph{116}, 015001\relax
\mciteBstWouldAddEndPuncttrue
\mciteSetBstMidEndSepPunct{\mcitedefaultmidpunct}
{\mcitedefaultendpunct}{\mcitedefaultseppunct}\relax
\EndOfBibitem
\bibitem[Fedeli \latin{et~al.}(2017)Fedeli, Sgattoni, Cantono, and
  Macchi]{fedeli2016}
Fedeli,~L.; Sgattoni,~A.; Cantono,~G.; Macchi,~A. Relativistic surface plasmon
  enhanced harmonic generation from gratings. \emph{Applied Physics Letters}
  \textbf{2017}, \emph{110}, 051103\relax
\mciteBstWouldAddEndPuncttrue
\mciteSetBstMidEndSepPunct{\mcitedefaultmidpunct}
{\mcitedefaultendpunct}{\mcitedefaultseppunct}\relax
\EndOfBibitem
\bibitem[Ye \latin{et~al.}(2014)Ye, Merlo, Burns, and Naughton]{ye2014}
Ye,~F.; Merlo,~J.~M.; Burns,~M.~J.; Naughton,~M.~J. Optical and electrical
  mappings of surface plasmon cavity modes. \emph{Nanophotonics} \textbf{2014},
  \emph{3}, 33--49\relax
\mciteBstWouldAddEndPuncttrue
\mciteSetBstMidEndSepPunct{\mcitedefaultmidpunct}
{\mcitedefaultendpunct}{\mcitedefaultseppunct}\relax
\EndOfBibitem
\bibitem[Maier(2007)]{Maier}
Maier,~S.~A. \emph{{Plasmonics: Fundamentals and Applications}}; Springer,
  2007\relax
\mciteBstWouldAddEndPuncttrue
\mciteSetBstMidEndSepPunct{\mcitedefaultmidpunct}
{\mcitedefaultendpunct}{\mcitedefaultseppunct}\relax
\EndOfBibitem
\bibitem[Quéré \latin{et~al.}(2014)Quéré, Vincenti, Borot, Monchocé,
  Hammond, Kim, Wheeler, Zhang, Ruchon, Auguste, Hergott, Villeneuve, Corkum,
  and Lopez-Martens]{Quere}
Quéré,~F.; Vincenti,~H.; Borot,~A.; Monchocé,~S.; Hammond,~T.~J.;
  Kim,~K.~T.; Wheeler,~J.~A.; Zhang,~C.; Ruchon,~T.; Auguste,~T.;
  Hergott,~J.~F.; Villeneuve,~D.~M.; Corkum,~P.~B.; Lopez-Martens,~R.
  Applications of ultrafast wavefront rotation in highly nonlinear optics.
  \emph{Journal of Physics B: Atomic, Molecular and Optical Physics}
  \textbf{2014}, \emph{47}, 124004\relax
\mciteBstWouldAddEndPuncttrue
\mciteSetBstMidEndSepPunct{\mcitedefaultmidpunct}
{\mcitedefaultendpunct}{\mcitedefaultseppunct}\relax
\EndOfBibitem
\bibitem[Wheeler \latin{et~al.}(2012)Wheeler, Borot, Monchoc{\'e}, Vincenti,
  Ricci, Malvache, Lopez-Martens, and Qu{\'e}r{\'e}]{wheeler2012}
Wheeler,~J.~A.; Borot,~A.; Monchoc{\'e},~S.; Vincenti,~H.; Ricci,~A.;
  Malvache,~A.; Lopez-Martens,~R.; Qu{\'e}r{\'e},~F. Attosecond lighthouses
  from plasma mirrors. \emph{Nature Photonics} \textbf{2012}, \emph{6},
  829--833\relax
\mciteBstWouldAddEndPuncttrue
\mciteSetBstMidEndSepPunct{\mcitedefaultmidpunct}
{\mcitedefaultendpunct}{\mcitedefaultseppunct}\relax
\EndOfBibitem
\bibitem[Vincenti and Qu\'er\'e(2012)Vincenti, and Qu\'er\'e]{Vincenti}
Vincenti,~H.; Qu\'er\'e,~F. Attosecond Lighthouses: How To Use Spatiotemporally
  Coupled Light Fields To Generate Isolated Attosecond Pulses. \emph{Physical
  Review Letters} \textbf{2012}, \emph{108}, 113904\relax
\mciteBstWouldAddEndPuncttrue
\mciteSetBstMidEndSepPunct{\mcitedefaultmidpunct}
{\mcitedefaultendpunct}{\mcitedefaultseppunct}\relax
\EndOfBibitem
\bibitem[Akturk \latin{et~al.}(2004)Akturk, Gu, Zeek, and Trebino]{Akturk}
Akturk,~S.; Gu,~X.; Zeek,~E.; Trebino,~R. Pulse-front tilt caused by spatial
  and temporal chirp. \emph{Opt. Express} \textbf{2004}, \emph{12},
  4399--4410\relax
\mciteBstWouldAddEndPuncttrue
\mciteSetBstMidEndSepPunct{\mcitedefaultmidpunct}
{\mcitedefaultendpunct}{\mcitedefaultseppunct}\relax
\EndOfBibitem
\bibitem[Smith \latin{et~al.}(2015)Smith, Stenger, Kristensen, Mortensen, and
  Bozhevolnyi]{smith2015}
Smith,~C.~L.; Stenger,~N.; Kristensen,~A.; Mortensen,~N.~A.; Bozhevolnyi,~S.~I.
  Gap and channeled plasmons in tapered grooves: a review. \emph{Nanoscale}
  \textbf{2015}, \emph{7}, 9355--9386\relax
\mciteBstWouldAddEndPuncttrue
\mciteSetBstMidEndSepPunct{\mcitedefaultmidpunct}
{\mcitedefaultendpunct}{\mcitedefaultseppunct}\relax
\EndOfBibitem
\bibitem[Gu \latin{et~al.}(2012)Gu, Ouyang, Jia, Stokes, Chen, Fahim, Li,
  Ventura, and Shi]{gu2012}
Gu,~M.; Ouyang,~Z.; Jia,~B.; Stokes,~N.; Chen,~X.; Fahim,~N.; Li,~X.;
  Ventura,~M.~J.; Shi,~Z. Nanoplasmonics: a frontier of photovoltaic solar
  cells. \emph{Nanophotonics} \textbf{2012}, \emph{1}, 235--248\relax
\mciteBstWouldAddEndPuncttrue
\mciteSetBstMidEndSepPunct{\mcitedefaultmidpunct}
{\mcitedefaultendpunct}{\mcitedefaultseppunct}\relax
\EndOfBibitem
\bibitem[Blaber \latin{et~al.}(2009)Blaber, Arnold, and Ford]{blaber2009}
Blaber,~M.~G.; Arnold,~M.~D.; Ford,~M.~J. Search for the ideal plasmonic
  nanoshell: the effects of surface scattering and alternatives to gold and
  silver. \emph{The Journal of Physical Chemistry C} \textbf{2009}, \emph{113},
  3041--3045\relax
\mciteBstWouldAddEndPuncttrue
\mciteSetBstMidEndSepPunct{\mcitedefaultmidpunct}
{\mcitedefaultendpunct}{\mcitedefaultseppunct}\relax
\EndOfBibitem
\bibitem[Lupetti \latin{et~al.}(2014)Lupetti, Hengster, Uphues, and
  Scrinzi]{lupetti2014}
Lupetti,~M.; Hengster,~J.; Uphues,~T.; Scrinzi,~A. Attosecond photoscopy of
  plasmonic excitations. \emph{Physical review letters} \textbf{2014},
  \emph{113}, 113903\relax
\mciteBstWouldAddEndPuncttrue
\mciteSetBstMidEndSepPunct{\mcitedefaultmidpunct}
{\mcitedefaultendpunct}{\mcitedefaultseppunct}\relax
\EndOfBibitem
\bibitem[Oskooi \latin{et~al.}(2009)Oskooi, Roundy, Ibanescu, Bermel,
  Joannopoulos, and Johnson]{Oskooi2009}
Oskooi,~A.~F.; Roundy,~D.; Ibanescu,~M.; Bermel,~P.; Joannopoulos,~J.~D.;
  Johnson,~S.~G. {Meep: A flexible free-software package for electromagnetic
  simulations by the FDTD method}. \emph{Computer Physics Communications}
  \textbf{2009}, \relax
\mciteBstWouldAddEndPunctfalse
\mciteSetBstMidEndSepPunct{\mcitedefaultmidpunct}
{}{\mcitedefaultseppunct}\relax
\EndOfBibitem
\bibitem[{Siegman}(1986)]{Siegman}
{Siegman},~A.~E. \emph{Lasers, by Anthony E.~Siegman.~Published by University
  Science Books, ISBN 0-935702-11-3, 1283pp, 1986.}; University Science Books,
  1986\relax
\mciteBstWouldAddEndPuncttrue
\mciteSetBstMidEndSepPunct{\mcitedefaultmidpunct}
{\mcitedefaultendpunct}{\mcitedefaultseppunct}\relax
\EndOfBibitem
\end{mcitethebibliography}

\providecommand{\latin}[1]{#1}
\providecommand*\mcitethebibliography{\thebibliography}
\csname @ifundefined\endcsname{endmcitethebibliography}
  {\let\endmcitethebibliography\endthebibliography}{}

\end{document}